\documentclass[aps,pre,twocolumn,showpacs,preprintnumbers,amsmath,amssymb]{revtex4}
\begin{document}
\title{Comment on 'Evolution of the unidirectional electromagnetic pulses in
an anisotropic two-level medium'}
\author{N.V. Ustinov}
 \email{n_ustinov@mail.ru}
 \noaffiliation
\date{\today}
\begin{abstract}
Recently, Zabolotskii [Phys. Rev. E 77, 036603 (2008)] presented the Lax pair 
for a version of the reduced Maxwell--Bloch equations. 
This version was derived under considering the unidirectional propagation of a 
two-component electromagnetic pulse through an anisotropic two-level medium. 
We demonstrate that his derivation contains essential omission, which led to 
the wrong version of the reduced Maxwell--Bloch equations. 
We also point out that the Lax pair for correct version of these equations had 
been known in the most general anisotropic case. 
\end{abstract}
\maketitle

A version of the reduced Maxwell--Bloch (RMB) equations was derived in 
Ref.~\cite{Z} as follows. 
The Hamiltonian of the interaction of two-component electromagnetic field with 
anisotropic two-level medium was taken in the form 
\begin{eqnarray}
\hat H&=&\frac{\hbar\omega_0}{2}\hat\sigma_3
-(d_{zx}^{(1)}E_x^{\prime}+d_{zy}^{(1)}E_y^{\prime})\hat\sigma_{11}
\nonumber\\&&
-(d_{zx}^{(2)}E_x^{\prime}+d_{zy}^{(2)}E_y^{\prime})\hat\sigma_{22}
-(d_{xx}E_x^{\prime}+d_{xy}E_y^{\prime})\hat\sigma_1
\nonumber\\&&
-(d_{yx}E_x^{\prime}+d_{yy}E_y^{\prime})\hat\sigma_2,
\label{H}
\end{eqnarray}
where $\hat\sigma_k$ ($k=1,2,3$) are the Pauli matrices, 
$\hat\sigma_{11}=\mbox{diag}(1,0)$, $\hat\sigma_{22}=\mbox{diag}(0,1)$, 
$E_x^{\prime}$ and $E_y^{\prime}$ are the transverse components of the 
electric field, $\omega_0$ is the frequency of the transition, $\hbar$ is the 
Plank's constant. 
Equivalent representation of $\hat H$ is 
\begin{eqnarray}
\hat H=\frac{\hbar\omega_0}{2}\hat\sigma_3-\hat d_xE_x^{\prime}-
\hat d_yE_y^{\prime}, 
\label{Ham}
\end{eqnarray}
where the matrices $\hat d_x$ and $\hat d_y$ corresponding to the operators of 
the dipole moment components are defined as given 
\begin{eqnarray}
&\hat d_x=\left(
\begin{array}{cc}
{d_{zx}^{(1)}}_{\mathstrut}&d_{xx}-id_{yx}\\
d_{xx}+id_{yx}&{d_{zx}^{(2)}}^{\mathstrut}
\end{array}
\right)_{\mathstrut},&
\nonumber\\
&\hat d_y=\left(^{\mathstrut}
\begin{array}{cc}
{d_{zy}^{(1)}}_{\mathstrut}&d_{xy}-id_{yy}\\
d_{xy}+id_{yy}&{d_{zy}^{(2)}}{\mathstrut}
\end{array}
\right).&
\nonumber
\end{eqnarray}
Coefficients $d_{zs}^{(k)}$ ($k=1,2$, $s=x,y$), $d_{sp}$ ($s,p=x,y$) of these 
matrices are assumed to be real constants. 

New variables were introduced by the transformation 
\begin{eqnarray}
\left(
\begin{array}{c}
{E_x}_{\mathstrut}^{\mathstrut}\\
{E_y}^{\mathstrut}_{\mathstrut}
\end{array}
\right)=T\left(
\begin{array}{c}
{E_x^{\prime}}_{\mathstrut}\\
{E_y^{\prime}}^{\mathstrut}
\end{array}
\right),
\label{EE}
\end{eqnarray}
where 
\[
T=\left(
\begin{array}{cc}
{\displaystyle\frac{d_{xx}}{\delta_x}}_{\mathstrut}^{\mathstrut}&
\displaystyle\frac{d_{xy}}{\delta_x}\\
{\displaystyle\frac{d_{yx}}{\delta_y}}^{\mathstrut}_{\mathstrut}&
\displaystyle\frac{d_{yy}}{\delta_y}
\end{array}
\right),
\]
\[
\delta_x=\sqrt{d_{xx}^2+d_{xy}^2},\quad\delta_y=\sqrt{d_{yx}^2+d_{yy}^2}.
\]
Then, Hamiltonian (\ref{H}) reads as 
\begin{eqnarray}
\hat H&=&\frac{\hbar\omega_0}{2}\hat\sigma_3
-(p_x^{(1)}E_x+p_y^{(1)}E_y)\hat\sigma_{11}\nonumber\\
&&-(p_x^{(2)}E_x+p_y^{(2)}E_y)\hat\sigma_{22}
-\delta_xE_x\hat\sigma_1-\delta_yE_y\hat\sigma_2.
\nonumber
\end{eqnarray}
Here 
\[
p_x^{(k)}=\delta_x\frac{d_{zx}^{(k)}d_{yy}-d_{zy}^{(k)}d_{yx}}{P_0},
\quad
p_y^{(k)}=\delta_y\frac{d_{zy}^{(k)}d_{xx}-d_{zx}^{(k)}d_{xy}}{P_0}
\]
($k=1,2$), $P_0=d_{xx}d_{yy}-d_{xy}d_{yx}$.

The dynamics of the two-level medium was described by the von~Neumann equation 
on the density matrix $\hat\rho$, 
\begin{eqnarray}
i\hbar\frac{\partial\hat\rho}{\partial t}=[\hat H,\hat\rho],
\label{rho_t}
\end{eqnarray}
with 
\[
\hat\rho=
\left(
\begin{array}{cc}
{\rho_{11}}_{\mathstrut}^{\mathstrut}&\rho_{12}\\
{\rho_{21}}^{\mathstrut}_{\mathstrut}&\rho_{22}
\end{array}
\right).
\label{rho}
\]
Eq.~(\ref{rho_t}) was rewritten in the terms of the components of the 
Bloch vector 
\[
S_x=\frac{\rho_{12}+\rho_{21}}{2},\quad
S_y=\frac{\rho_{12}-\rho_{21}}{2i},\quad
S_z=\frac{\rho_{11}-\rho_{22}}{2}
\]
and  the dimensionless variables 
\[
{\cal E}_x=\frac{d_0E_x}{\hbar\omega_0},\quad
{\cal E}_y=\frac{d_0E_y}{\hbar\omega_0},
\]
where 
\[
d_0=\sqrt{4\delta_x^2+4\delta_y^2+\left(p_x^{(1)}-p_x^{(2)}\right)^2+
\left(p_y^{(1)}-p_y^{(2)}\right)^2},
\]
as given 
\begin{eqnarray}
&\displaystyle{\frac{\partial S_x}{\partial\tau^{\prime}}}_{\mathstrut}
=(1-m_x{\cal E}_x-m_y{\cal E}_y)S_y+\mu_y{\cal E}_yS_z,&
\label{S_x_t}\\
&\displaystyle{\frac{\partial S_y}{\partial\tau^{\prime}}}_{\mathstrut}
=(m_x{\cal E}_x+m_y{\cal E}_y-1)S_x-\mu_x{\cal E}_xS_z,&
\label{S_y_t}\\
&\displaystyle\frac{\partial S_z}{\partial\tau^{\prime}}=\mu_x{\cal E}_xS_y-
\mu_y{\cal E}_yS_x.&
\label{S_z_t}
\end{eqnarray}
Here $\tau^{\prime}=\omega_0t$, 
\begin{eqnarray}
&\displaystyle\mu_x=\frac{2\delta_x}{d_0},\quad 
m_x=\frac{p_x^{(1)}-p_x^{(2)}}{d_0}_{\mathstrut},&
\nonumber\\
&\displaystyle\mu_y=-\frac{2\delta_y}{d_0},\quad 
m_y=\frac{p_y^{(1)}-p_y^{(2)}}{d_0}^{\mathstrut}.&
\nonumber
\end{eqnarray}

An evolution of the electromagnetic field of the pulse has to obey the Maxwell 
equations if the semiclassical approach is applied. 
It was claimed in \cite{Z} that the transformed components $E_x$ and $E_y$ 
satisfy the ''Maxwell equations'' 
\begin{eqnarray}
&\displaystyle{\frac{\partial^2 E_x}{\partial z^2}}_{\mathstrut}
-\frac{n^2}{c^2}\frac{\partial^2 E_x}{\partial t^2}=
\frac{4\pi}{c^2}\frac{\partial^2P_x}
{\partial t^2},&
\label{E_x_zz}\\
&\displaystyle{\frac{\partial^2 E_y}{\partial z^2}}^{\mathstrut}
-\frac{n^2}{c^2}\frac{\partial^2 E_y}{\partial t^2}=
\frac{4\pi}{c^2}\frac{\partial^2P_y}
{\partial t^2},&
\label{E_y_zz}
\end{eqnarray}
where $n$ is the refractive index of the medium, $c$ is the light speed in 
free space. 
The quantities $P_x$ and $P_y$ in the right hand sides of Eqs.~(\ref{E_x_zz}) 
and (\ref{E_y_zz}) were interpreted as the components of the medium 
polarization and were defined for this reason in the following manner: 
\begin{eqnarray}
&\displaystyle P_x=-N\,\mbox{Tr}\left\{\hat\rho\frac{\partial\hat H}
{\partial E_x}\right\}_{\mathstrut},&
\label{P_x}\\
&\displaystyle P_y=-N\,\mbox{Tr}\left\{\hat\rho\frac{\partial\hat H}
{\partial E_y}\right\},&
\label{P_y}
\end{eqnarray}
where $N$ is the density of the medium. 

At last, the reduced equations 
\begin{eqnarray}
&\displaystyle{\frac{\partial{\cal E}_x}{\partial\chi^{\prime}}}_{\mathstrut}
=R_z{\cal E}_y-\mu_xS_y,&
\label{E_x_z}\\
&\displaystyle\frac{\partial{\cal E}_y}{\partial\chi^{\prime}}=
-R_z{\cal E}_x+\mu_yS_x,&
\label{E_y_z}
\end{eqnarray}
where 
\begin{eqnarray}
&\displaystyle\chi^{\prime}=\frac{2\pi Nd_0^2}{n\hbar c}\left(z+\frac{c}{n}t
\right)_{\mathstrut},&
\nonumber\\
&R_z=m_x\mu_yS_x+m_y\mu_xS_y-\mu_x\mu_yS_z^{\mathstrut},&
\nonumber
\end{eqnarray}
were obtained from (\ref{E_x_zz}) and (\ref{E_y_zz}) with the help of the 
unidirectional propagation approximation. 

The misprints were corrected in the formulas presented above. 
We divided the first term in the right hand side of (\ref{H}) by two, changed 
a sign in the definition of $\mu_y$, multiplied variable $\chi^{\prime}$ 
by $\omega_0/2$. 
Also, Eqs.~(\ref{E_x_zz})--(\ref{P_y}) were written in the terms of the 
variables $E_x$, $E_y$ instead of ${\cal E}_x$, ${\cal E}_y$, and the 
multipliers $d_x$, $d_y$ are omitted in the right hand sides of 
Eqs.~(\ref{E_x_zz}), (\ref{E_y_zz}). 
We found that these corrections are necessary for the system 
(\ref{S_x_t})--(\ref{S_z_t}), (\ref{E_x_z}), (\ref{E_y_z}) to be obtained. 

According to Ref.~\cite{Z}, an evolution of the unidirectional two-component 
electromagnetic pulses in an anisotropic two-level medium is described by the 
version (\ref{S_x_t})--(\ref{S_z_t}), (\ref{E_x_z}), (\ref{E_y_z}) of the RMB 
equations. 
We believe this statement to be misleading since the derivation of 
Eqs.~(\ref{E_x_z}), (\ref{E_y_z}) contains essential omission. 

To explain why this statement in \cite{Z} is incorrect we consider the Maxwell 
equations for the components $E_x^{\prime}$ and $E_y^{\prime}$ of the electric 
field. 
So, we have 
\begin{eqnarray}
&\displaystyle{\frac{\partial^2 E_x^{\prime}}{\partial z^2}}_{\mathstrut}
-\frac{n^2}{c^2}\frac{\partial^2 E_x^{\prime}}{\partial t^2}=\frac{4\pi}{c^2}
\frac{\partial^2P_x^{\prime}}{\partial t^2},&
\label{E_x_'_zz}\\
&\displaystyle{\frac{\partial^2 E_y^{\prime}}{\partial z^2}}^{\mathstrut}
-\frac{n^2}{c^2}\frac{\partial^2 E_y^{\prime}}{\partial t^2}=\frac{4\pi}{c^2}
\frac{\partial^2P_y^{\prime}}{\partial t^2},&
\label{E_y_'_zz}
\end{eqnarray}
where the components of the medium polarization are 
\begin{eqnarray}
&\displaystyle P_x^{\prime}=-N\,\mbox{Tr}\left\{\hat\rho\frac{\partial\hat H}
{\partial E_x^{\prime}}\right\}_{\mathstrut},&
\label{P_x_'}\\
&\displaystyle P_y^{\prime}=-N\,\mbox{Tr}\left\{\hat\rho\frac{\partial\hat H}
{\partial E_y^{\prime}}\right\}&
\label{P_y_'}
\end{eqnarray}
(compare with Eqs.~(\ref{P_x}), (\ref{P_y})). 
Substitution of (\ref{Ham}) into (\ref{P_x_'}), (\ref{P_y_'}) leads to the 
standard formulas: 
\[
P_x^{\prime}=N\,\mbox{Tr}\,(\hat\rho\hat d_x),\quad
P_y^{\prime}=N\,\mbox{Tr}\,(\hat\rho\hat d_y).
\]
From Eqs.~(\ref{E_x_'_zz}), (\ref{E_y_'_zz}) and the transformation 
(\ref{EE}), one obtains the wave equations on transformed components $E_x$ and 
$E_y$: 
\begin{eqnarray}
&\displaystyle{\frac{\partial^2 E_x}{\partial z^2}}_{\mathstrut}
-\frac{n^2}{c^2}\frac{\partial^2 E_x}{\partial t^2}=\frac{4\pi}{c^2}
\frac{\partial^2\tilde P_x}{\partial t^2},&
\label{t_E_x_zz}\\
&\displaystyle{\frac{\partial^2 E_y}{\partial z^2}}^{\mathstrut}
-\frac{n^2}{c^2}\frac{\partial^2 E_y}{\partial t^2}=\frac{4\pi}{c^2}
\frac{\partial^2\tilde P_y}{\partial t^2},&
\label{t_E_y_zz}
\end{eqnarray}
where quantities $\tilde P_x$ and $\tilde P_y$ are defined by the relation 
\begin{eqnarray}
\left(
\begin{array}{c}
\mbox{$\tilde P_x$}_{\mathstrut}\\
\mbox{$\tilde P_y$}_{\mathstrut}^{\mathstrut}
\end{array}
\right)=T\left(
\begin{array}{c}
{P_x^{\prime}}_{\mathstrut}\\
{P_y^{\prime}}^{\mathstrut}
\end{array}
\right).
\label{tPP'}
\end{eqnarray}

A connection between the components $P_x^{\prime}$, $P_y^{\prime}$ and $P_x$, 
$P_y$ exists also. 
Taking into account (\ref{EE}), we deduce from Eqs.~(\ref{P_x}), (\ref{P_y}) 
and (\ref{P_x_'}), (\ref{P_y_'}) that 
\begin{eqnarray}
\left(
\begin{array}{c}
{P_x^{\prime}}_{\mathstrut}\\
{P_y^{\prime}}^{\mathstrut}
\end{array}
\right)=T^T\left(
\begin{array}{c}
{P_x}_{\mathstrut}^{\mathstrut}\\
{P_y}^{\mathstrut}_{\mathstrut}
\end{array}
\right).
\label{P'P}
\end{eqnarray}

Define the dimensionless parameter 
\[
\varepsilon=\frac{d_{xx}d_{yx}+d_{xy}d_{yy}}{\delta_x\delta_y}. 
\]
If the condition 
\begin{eqnarray}
\varepsilon=0
\label{cond}
\end{eqnarray}
holds, then matrix $T$ is orthogonal: $T^T=T^{-1}$. 
It can easily be seen from Eqs.~(\ref{tPP'}) and (\ref{P'P}) that 
$\tilde P_x=P_x$, $\tilde P_y=P_y$ in this case. 
Formula (\ref{EE}) under this condition is nothing but the rotation 
transformation. 

In the general case ($\varepsilon\ne0$), we have from (\ref{tPP'}), 
(\ref{P'P}) that $\tilde P_x\ne P_x$ and $\tilde P_y\ne P_y$, i.e. the right 
hand sides of the equations (\ref{E_x_zz}), (\ref{E_y_zz}) and 
(\ref{t_E_x_zz}), (\ref{t_E_y_zz}) on transformed components $E_x$, $E_y$ are 
different. 
This shows that it is incorrect to determine the quantities $P_x$ and $P_y$ in 
Eqs.~(\ref{E_x_zz}), (\ref{E_y_zz}) by means of the formulas (\ref{P_x}) and 
(\ref{P_y}) if $\varepsilon\ne0$. 
These formulas are valid in the particular case $\varepsilon=0$ when 
transformation (\ref{EE}) is the rotation transformation. 
Thus, the version (\ref{S_x_t})--(\ref{S_z_t}), (\ref{E_x_z}), (\ref{E_y_z}) 
of the RMB equations can be applied only if the condition (\ref{cond}) is 
imposed on the elements of the matrices $\hat d_x$ and $\hat d_y$. 
It is wrong to exploit this version in the general case. 

An application of the unidirectional propagation approximation to 
Eqs.~(\ref{t_E_x_zz}) and (\ref{t_E_y_zz}) gives 
\begin{eqnarray}
&\displaystyle{\frac{\partial{\cal E}_x}{\partial\chi^{\prime}}}_{\mathstrut}
=R_z{\cal E}_y-\mu_xS_y-\varepsilon(R_z{\cal E}_x-\mu_yS_x),&
\label{t_E_x_z}\\
&\displaystyle\frac{\partial{\cal E}_y}{\partial\chi^{\prime}}=
-R_z{\cal E}_x+\mu_yS_x+\varepsilon(R_z{\cal E}_y-\mu_xS_y).&
\label{t_E_y_z}
\end{eqnarray}

The wrong version (\ref{S_x_t})--(\ref{S_z_t}), (\ref{E_x_z}), (\ref{E_y_z}) 
of the RMB equations contains four parameters $\mu_x$, $\mu_y$, $m_x$ and 
$m_y$ connected by the relation 
\begin{eqnarray}
\mu_x^2+\mu_y^2+m_x^2+m_y^2=1.
\label{rel}
\end{eqnarray}
It was claimed in \cite{Z} that the Lax pair exists for this version. 

The correct version (\ref{S_x_t})--(\ref{S_z_t}), (\ref{t_E_x_z}), 
(\ref{t_E_y_z}) of the RMB equations contains five parameters $\mu_x$, 
$\mu_y$, $m_x$, $m_y$ and $\varepsilon$. 
The relation (\ref{rel}) is fulfilled also. 
The system of the RMB equations equivalent to the correct version was 
considered in Ref.~\cite{U}. 
It was shown that this system possesses the Lax pair in the most general 
anisotropic case when all the elements of the matrices $\hat d_x$ and 
$\hat d_y$ are arbitrary. 

The system of the RMB equations and its Lax pair were written in \cite{U} in 
the terms of physical variables and parameters. 
Having rewritten these systems in the terms of the variables 
$\tau^{\prime}$, $\chi^{\prime}$ and ${\cal E}_x$, ${\cal E}_y$, we obtain 
equations (\ref{S_x_t})--(\ref{S_z_t}), (\ref{t_E_x_z}), (\ref{t_E_y_z}) and 
their Lax pair: 
\begin{eqnarray}
&\displaystyle{\frac{\partial\psi}{\partial\tau^{\prime}}}_{\mathstrut}
=L(\lambda)\psi,&
\label{Lax_1}\\
&\displaystyle\frac{\partial\psi}{\partial\chi^{\prime}}=A(\lambda)\psi.&
\label{Lax_2}
\end{eqnarray}
Here $\psi=\psi(\tau^{\prime},\chi^{\prime},\lambda)$ is a solution of the Lax 
pair, $\lambda$ is the spectral parameter, $2\times2$ matrices $L(\lambda)$ 
and $A(\lambda)$ are defined as given 
\begin{eqnarray}
&\displaystyle L(\lambda)=\frac12\left( 
\begin{array}{cc}
\displaystyle i\left[\lambda^2-\frac{b}{\lambda^2}\right]_{\mathstrut}&
\displaystyle\lambda E^*+\frac{\delta_2}{\delta_1}\frac{E}{\lambda}\\
\displaystyle\lambda\delta_1 E+\delta_2^*\frac{E^*}{\lambda}&
\displaystyle-i\left[\lambda^2-\frac{b}{\lambda^2}\right]^{\mathstrut}
\end{array}
\right)_{\mathstrut},&
\label{L}\\
&A(\lambda)=r\left(
\begin{array}{cc}
\displaystyle-i\left[\lambda^2-\frac{b}{\lambda^2}\right]_{\mathstrut}R_z&
\displaystyle\frac{\mu_x\mu_y}{\delta_1}R\\     
\displaystyle\mu_x\mu_yR^*&
\displaystyle i\left[\lambda^2-\frac{b}{\lambda^2}\right]^{\mathstrut}R_z 
\end{array}
\right)_{\mathstrut},\mbox{\ }&
\label{A}\\
&\displaystyle{E=\frac{E_x-s^*E_y}{\sqrt{1-\varepsilon^2}}+
\frac{\delta_3}{\delta_1}}_{\mathstrut},\quad 
R=\lambda Q^*+\frac{\delta_2}{\delta_1}\frac{Q}{\lambda},
\nonumber\\
&{Q=\delta_3S_z+\delta_4S_x+\delta_5S_y}_{\mathstrut},&
\nonumber\\
&\displaystyle{\delta_1=-\sqrt{1-\varepsilon^2}\,
\frac{\mu_x^2\mu_y^2+\mu_x^2m_y^2+\mu_y^2m_x^2}{\mu_x\mu_y}}_{\mathstrut},& 
\nonumber\\
&\displaystyle{\delta_2=\frac{s^2(\mu_x^2+m_x^2)+\mu_y^2+m_y^2+2sm_xm_y}{4}
}_{\mathstrut},&
\nonumber\\
&\displaystyle\delta_3=\frac{\mu_y^2m_x-s^*\mu_x^2m_y}{\mu_x\mu_y},\quad 
{\delta_4=\frac{\mu_y^2+m_y^2+s^*m_xm_y}{\mu_y}}_{\mathstrut},& 
\nonumber\\
&\displaystyle{\delta_5=-\frac{s^*(\mu_x^2+m_x^2)+m_xm_y}{\mu_x}
}_{\mathstrut},& 
\nonumber\\
&\displaystyle{b=\frac{|\delta_2|^2}{\delta_1^2}}_{\mathstrut},\quad 
s=\varepsilon+i\sqrt{1-\varepsilon^2},&
\nonumber\\
&\displaystyle r=-\frac12\frac{\sqrt{1-\varepsilon^2}}
{\displaystyle\lambda^2+\frac{b}{\lambda^2}+
\frac{1+2\varepsilon m_xm_y}{2\delta_1}}.& 
\nonumber
\end{eqnarray}

It can be checked immediately that the overdetermined system (\ref{Lax_1}), 
(\ref{Lax_2}) is the Lax pair of the correct version of the RMB equations 
(\ref{S_x_t})--(\ref{S_z_t}), (\ref{t_E_x_z}), (\ref{t_E_y_z}). 
Indeed, the compatibility condition of Eqs.~(\ref{Lax_1}), (\ref{Lax_2}) is 
\begin{eqnarray}
\frac{\partial L(\lambda)}{\partial\chi^{\prime}}-
\frac{\partial A(\lambda)}{\partial\tau^{\prime}}+[L(\lambda),A(\lambda)]=0. 
\label{cc}
\end{eqnarray}
A substitution of (\ref{L}) and (\ref{A}) into (\ref{cc}) yields the system of 
the RMB equations (\ref{S_x_t})--(\ref{S_z_t}), (\ref{t_E_x_z}), 
(\ref{t_E_y_z}). 

The system equivalent to Eqs.~(\ref{S_x_t})--(\ref{S_z_t}), (\ref{t_E_x_z}), 
(\ref{t_E_y_z}) is obtained by applying the unidirectional propagation 
approximation to the Maxwell equations (\ref{E_x_'_zz}), (\ref{E_y_'_zz}). 
The coefficients of the Lax pair found in \cite{U} for this system were 
expressed directly through the elements of the matrices $\hat d_x$ and 
$\hat d_y$.

\end{document}